**Anatoly Bessalov**
DSc, professor
Borys Grinchenko Kyiv University, Kyiv, Ukraine
ORCID: 0000-0002-6967-5001
*a.bessalov@kubg.edu.ua*

**Evgeniy Grubiyan**
MSc
National Technical University of Ukraine "Igor Sikorsky Kyiv Polytechnic Institute", Kyiv, Ukraine
ORCID: 0000-0002-6967-5001
*grubian.euhen@gmail.com*

**Volodymyr Sokolov**
PhD
Borys Grinchenko Kyiv University, Kyiv, Ukraine
ORCID: 0000-0002-9349-7946
*v.sokolov@kubg.edu.ua*

**Pavlo Skladannyi**
Senior lecturer
Borys Grinchenko Kyiv University, Kyiv, Ukraine
ORCID: 0000-0002-7775-6039
*p.skladannyi@kubg.edu.ua*

## 3- AND 5-ISOGENIES OF SUPERSINGULAR EDWARDS CURVES

**Abstract.** An analysis is made of the properties and conditions for the existence of 3- and 5-isogenies of complete and quadratic supersingular Edwards curves. For the encapsulation of keys based on the SIDH algorithm, it is proposed to use isogeny of minimal odd degrees 3 and 5, which allows bypassing the problem of singular points of the 2nd and 4th orders, characteristic of 2-isogenies. A review of the main properties of the classes of complete, quadratic, and twisted Edwards curves over a simple field is given. Equations for the isogeny of odd degrees are reduced to a form adapted to curves in the form of Weierstrass. To do this, use the modified law of addition of curve points in the generalized Edwards form, which preserves the horizontal symmetry of the curve return points. Examples of the calculation of 3- and 5-isogenies of complete Edwards supersingular curves over small simple fields are given, and the properties of the isogeny composition for their calculation with large-order kernels are discussed. Equations are obtained for upper complexity estimates for computing isogeny of odd degrees 3 and 5 in the classes of complete and quadratic Edwards curves in projective coordinates; algorithms are constructed for calculating 3- and 5-isogenies of Edwards curves with complexity $6M + 4S$ and $12M + 5S$, respectively. The conditions for the existence of supersingular complete and quadratic Edwards curves of order $4 \cdot 3^m \cdot 5^n$ and $8 \cdot 3^m \cdot 5^n$ are found. Some parameters of the cryptosystem are determined when implementing the SIDH algorithm at the level of quantum security of 128 bits.

## 1. INTRODUCTION

One of the well-known prospects of post-quantum cryptography (PQC) is the algorithms based on the isogeny of supersingular elliptic curves with as many subgroups of their points as possible (in particular, the SIDH algorithm [1]). The problem of the discrete logarithm of classical elliptic cryptography is replaced by the problem of finding one of the isogenous sets of subgroups of such a non-cyclic curve that is sufficiently resistant to the attacks of a quantum computer. To date, the growing interest in isogeny is associated with the shortest key length in the proposed algorithms in comparison with other well-known candidates for post-quantum cryptography at a given level of strength.

This article deals with the properties of 3- and 5-isogenies of two classes of these curves, in particular the conditions of their existence. Section 2 provides a brief review of the literature on this topic. In Section 3, we touch upon the issue of how to solve the problem of singular points that occurs when programming the SIDH algorithm on Edwards curves using 2-isogenies. Instead of 2- and 3-isogenies, it is proposed to construct an algorithm on 3- and 5-isogenies of points of odd orders, which allows circumventing singular points. Section 4 gives a brief overview of the properties of three classes of Edwards curves according to the new classification. In Section 5, we prove a equation for the isogeny of odd degrees expressed by rational functions of one variable and give examples. In Section 6, estimates are obtained for the complexity of computing 3- and 5-isogenies in projective coordinates, taking into account operations with kernel coordinates. Finally, in Section 7, the conditions for the existence of 3- and 5-isogenies and the requirements for the curve parameters for the SIDH algorithm are defined [1].

## 2. REVIEW OF LITERATURE

The properties of isogeny for curves in the form of Weierstrass are sufficiently studied. Effective construction methods and isogeny properties of promising classes of curves in the Edwards form are less known.

Edwards curves with one parameter, defined in [2], have very attractive advantages for cryptography: maximum point exponentiation speed, completeness, and universality of the point addition law, affine coordinates of a neutral element of a group of points, increased security against side-channel attacks. The programming of group operations becomes more efficient and accelerates due to the absence of a singular point at infinity as the zero of an abelian group of points. The introduction of the second parameter of the curve in [3] expanded the class of curves in the Edwards form and generated curves with new properties that are interesting for cryptographic applications.

Along with the properties noted above, curves in the Edwards form proved to be the fastest technology in calculating isogeny. In [4], experimental estimates of the rate of calculation of isogeny on Edwards curves are presented, more than three times higher than the indices for curves in the Weierstrass form. Since the procedure for finding an isogenic point usually includes the scalar product of the point, the complex gain in the speed of the algorithms on the Edwards curves becomes significant.

## 3. STATEMENT OF THE PROBLEM

Well-known implementations of the SIDH algorithm mainly use curves in the forms of Weierstrass and Montgomery. Our attempt to programmatically implement the SIDH algorithm using 2- and 3-isogenies of curves in the Edwards form encountered the problem of the presence of 4 singular points at infinity of the 2nd and 4th orders in the class of quadratic Edwards curves, to which all Edwards curves are mapped over the field $F_{p^2}$ set over the field $F_p$. These points exist in all subgroups of even orders, the number of which exceeds half of all subgroups of the curve. The appearance of any singular point in the calculation of isogeny significantly slows down the software implementation of the SIDH algorithm on Edwards curves. To get around this problem, we propose using isogenies of minimal odd degrees 3 and 5 for points of odd order of the curve. Although the transition from 2- to 5-isogeny complicates the calculation algorithm, such a smooth implementation of the algorithm is faster.

Among the numerous works on this problem, we single out articles [4, 5], in which isogeny equations for curves in the Edwards form were first obtained. Our analysis in this paper is based on their results using the properties of supersingular curves [6]. To adapt the definitions for the arithmetic of isogeny of Edwards curves and curves in the Weierstrass form, we use the modified law of addition of points [7, 9].

## 4. CLASSES OF CURVES IN THE GENERALIZED EDWARDS FORM

The elliptic curve in the generalized Edwards form [3,7,8] is determined by the equation

$$E_{a,d}: x^2 + ay^2 = 1 + dx^2y^2, a, d \in F_p^*, d \neq 1, a \neq d, d \neq 2. \tag{1}$$

In contrast to the equation of this curve in [3], here we multiply the parameter $a$ by $y^2$ instead of $x^2$. If it is quadratic $\chi(ad) = -1$, the curve (1) is isomorphic to the *complete Edwards curve* [1] with one parameter $d$.

$$E_d: x^2 + y^2 = 1 + dx^2y^2, \chi(d) = -1, d \neq 0,1. \tag{2}$$

In case $\chi(ad) = 1$ and $\chi(a) = \chi(d) = 1$, then there is an isomorphism of the curve (1) with a *quadratic Edwards curve* [7]

$$E_d: x^2 + y^2 = 1 + dx^2y^2, \chi(d) = 1, d \neq 0,1, \tag{3}$$

having, in contrast to (2), the parameter $d$ defined as a square. This difference leads to radically different properties of the curves (2) and (3) [7], which are summarized below. Despite this, in the world literature, these classes of curves are united by the common term *Edwards curves* [3].

Curves with different values $d$ are isomorphic if they have the same *j*-invariant equal to the curve (1)

$$j(a, d) = \frac{16(a^2 + d^2 + 14ad)}{ad(a - d)^4}.$$

This parameter is basic in the structure of graphs of isogenic curves, the vertices of which define classes of isomorphic curves.

In our article [9], we proposed interchanging the coordinates x and y in the form of the Edwards curve. Then the modified universal law of addition of points of the curve (1) has the form:

$$(x_1, y_1) + (x_2, y_2) = \left( \frac{x_1x_2 - ay_1y_2}{1 - dx_1x_2y_1y_2}, \frac{x_1y_2 + x_2y_1}{1 + dx_1x_2y_1y_2} \right). \tag{4}$$

If two points coincide, we obtain from (2) the law of doubling points

$$2(x_1, y_1) = \left( \frac{x_1^2 - ay_1^2}{1 - dx_1^2y_1^2}, \frac{2x_1y_1}{1 + dx_1^2y_1^2} \right). \tag{5}$$

Using the modified laws (4), (5) allows us to preserve the generally accepted horizontal symmetry (relative to the axis $x$) of the reverse points. Defining now the reverse point as $-P = (x_1, -y_1)$ we obtain, according to (4), the coordinates of the neutral element of the group of points $O = (x_1, y_1) + (x_1, -y_1) = (1,0)$. In addition to the neutral element $O$, the axis $X$ also always contains the second-order point $D_0 = (-1,0)$, for which, by (5) $D_0 = (-1,0), 2D_0 = (1,0) = O$. Depending on the properties of the parameters $a$ and $d$, you can get two more singular points of the $2^{nd}$ order and two or more points of the $4^{th}$ order. As follows from (1), points $\pm D_0 = \left(0, \pm \frac{1}{\sqrt{a}}\right)$ of the fourth-order may lie on the axis y, for which $\pm 2F_0 = D_0 = (-1,0)$. These points exist over the simple field $F_p$, if the parameter is a square (quadratic residue). From equation (1) we define the squares:

$$x^2 = \frac{1 - ay^2}{1 - dy^2}, y^2 = \frac{1 - x^2}{a - dx^2},$$

generating singular points at infinity (we put the sign $\infty$ when dividing by 0):

$$D_{1,2} = \left(\pm \sqrt{\frac{a}{d}}, \infty\right), \pm F_{11} = \left(\infty, \pm \frac{1}{\sqrt{d}}\right). \tag{6}$$

They arise in cases $\chi(ad) = 1$ and $\chi(d) = 1$ respectively. This, for example, is always performed in the extension of the field $F_{p^2}$. According to the rules of terminal transition and the doubling law (5), we can verify that $2D_{1,2} = O, \pm 2F_1 = D_0 = (-1,0)$. In other words, under the conditions of their existence, singular points $D_{1,2}$ are points of the $2^{nd}$ order, and singular points $\pm F_1$ are points of the $4^{th}$ order.

Depending on the properties of the parameters $a$ and $d$, the curves in the generalized Edwards form (1) is divided into 3 disjoint (non-isomorphic) classes [7]:

- complete Edwards curves with the condition C1: $\chi(ad) = -1$;
- twisted Edwards curves with the condition C2.1: $\chi(a) = \chi(d) = -1$;
- quadratic Edwards curves with the condition C2.2: $\chi(a) = \chi(d) = 1$.

The main properties of these classes of curves [7, 9]:

1. Concerning the points of the second order, the first class of complete Edwards curves over a simple field is the class of *cyclic* curves (with one point of the second order), twisted and quadratic Edwards curves form classes of *non-cyclic curves* (3 points of the second-order each). The maximum order of the points of the curves of the last classes does not exceed $N_E/2$.

2. The class of complete Edwards curves does not contain singular points.

3. Twisted Edwards curves contain only two second-order singular points $D_{1,2} = \left(\pm \sqrt{\frac{a}{d}}; \infty\right)$, and Edwards quadratic curves, in addition to them,—two other fourth-order singular points $\pm F_{11} = \left(\infty; \pm \frac{1}{\sqrt{d}}\right)$.

4. Edwards twisted and quadratic curves form quadratic torsion pairs based on a parameter transformation: $\tilde{a} = ca, \tilde{d} = cd, \chi(c) = -1$.

5. In the classes of twisted and quadratic Edwards curves, the replacement $a \leftrightarrow d$ gives the isomorphism $E_{a,d} \sim E_{d,a}$.

6. Complete and quadratic Edwards curves are isomorphic to the curves with parameter $a = 1: E_{a,d} \sim E_{1,d/a}$. The introduction of the new parameter into the equation of curve (1) is justified only for the class of twisted Edwards curves.

7. The twisted Edwards curves $p \equiv 1 \bmod 4$ do not have the $4^{th}$ order points.

We emphasize that in the extension $F_{p^2}$ of the simple field $F_p$ all 3 classes of Edwards curves defined over a simple field acquire the properties of quadratic curves (3). Therefore, further, we consider mainly curves $E_d$ of the form (2) and (3).

## 5. ISOGENIES OF ODD DEGREES OF EDWARDS CURVES

The isogeny of the elliptic curve $E(K)$ over the field K into the curve $E'(K)$ is the homomorphism $\phi : E(\bar{K}) \to E'(\bar{K})$, defined by rational functions. That means that for all $P, Q \in E(K)$ $\phi(P+Q) = \phi(P) + \phi(Q)$ there exists the rational function [10]

$$\phi(x, y) = \left( \frac{p(x)}{q(x)}, y \frac{f(x)}{g(x)} \right) = (x', y'),\qquad(7)$$

mapping curve points E to curve points $E'$. The degree of isogeny is called the maximum of the degrees $l = \deg \phi(x, y) = \max\{\deg p(x), \deg q(x)\}$ and its kernel is $\ker \phi = G$ for subgroup $G \subseteq E$, the points of which are reflected by the function $\phi(x, y)$ into the neutral element $O$ of the group $E'$. The degree of separable isogeny is equal to the order $l$ of its kernel. Isogeny compresses the curve points $E$ $l$ times ($l$ curve points are displayed at one point of the curve $E'$). At $G = O$ isogeny becomes the isomorphism with the degree 1.

The basis of the construction of isogeny of odd simple degrees for Edwards curves is based on Theorem 2 [4].

Let's formulate it taking into account the modification (4) of the law of addition of points of the curve (1) at $a = 1$.

**Theorem 2 [4].** Let's $G = \{(1,0), \pm Q_1, \pm Q_2, ..., \pm Q_s\}$.

The subgroup of the odd order $l = 2s + 1$ of points $\pm Q_i = (\alpha_i, \pm \beta_i)$ of the curve $E_d$.

Let's define

$$\phi(P) = \left( \prod_{Q \in G} \frac{x_{P+Q}}{x_Q}, \prod_{Q \in G} \frac{y_{P+Q}}{x_Q} \right)$$

Then $\phi(x, y)$ is the $l$-isogeny with the kernel $G$ from the curve $E_d$ into the curve $E_{d'}'$ with the parameter $d' = A^8 d^l$, $A = \prod_{i=1}^{s} \alpha_i$, and the mapping function

$$\phi(x, y) = \left( \frac{x}{A^2} \prod_{i=1}^{s} \frac{(\alpha_i x)^2 - (\beta_i y)^2}{1 - (d\alpha_i \beta_i xy)^2}, \frac{y}{A^2} \prod_{i=1}^{s} \frac{(\alpha_i y)^2 - (\beta_i x)^2}{1 - (d\alpha_i \beta_i xy)^2} \right).\qquad(8)$$

Its proof is given in [4]. An important consequence of this is that isogenic curves lie in the same classes as curves $E_d$ (i. e., complete Edwards curves are mapped to complete and quadratic curves—to quadratic). This significantly distinguishes the isogeny of odd degrees from the 2-isogeny (for them, the complete Edwards curves are mapped into quadratic ones).

The equation (8) for the function $\phi(x, y)$ directly follows the definition $\phi(P)$ in the statement of the theorem, the equation (4) of the addition of points $(x_P, y_P) = (x, y)$ with the points $\pm Q_i = (\alpha_i, \pm \beta_i)$, wherein, for pairs of coordinates we have

$$\frac{x_{P+Q_i}}{x_{Q_i}}\frac{x_{P-Q_i}}{x_{-Q_i}} = \frac{1}{\alpha_i^2}\frac{(\alpha_i x)^2 - (\beta_i y)^2}{1-(d\alpha_i\beta_i xy)^2},$$

$$\frac{y_{P+Q_i}}{x_{Q_i}}\frac{y_{P-Q_i}}{x_{-Q_i}} = \frac{1}{\alpha_i^2}\frac{(\beta_i x)^2 - (\alpha_i y)^2}{1-(d\alpha_i\beta_i xy)^2}.$$

The multipliers x and y before the products in the coordinates of the function $\phi(x,y)$ take into account the neutral element $O = (1_i, 0)$ of the kernel of isogeny.

From (8) the property $\phi(1,0) = (1,0)$ it is obvious that i. e. the neutral element is mapped in itself. For all points of the kernel $\phi(\pm Q_i = (\alpha_i, \pm\beta_i)) = (1,0)$ is also true.

The mapping (8) can be reduced to the form (7), then the determination of the degree of isogeny becomes obvious. From (2) and (3) let's express $y^2 = (1-x^2)/(1-dx^2)$, and substitute this value in (8). Then in the numerator of the first coordinate (8)

$$\alpha_i^2 x^2 - \beta_i^2 y^2 = \alpha_i^2 x^2 - \beta_i^2\frac{1-x^2}{1-dx^2} = \frac{(\alpha_i^2+\beta_i^2)x^2 - \beta_i^2 - d\alpha_i^2 x^4}{1-dx^2} = \frac{(1+d\alpha_i^2\beta_i^2)x^2 - \beta_i^2 - d\alpha_i^2 x^4}{1-dx^2} =$$

$$= \frac{x^2 - \beta_i^2 - d(\alpha_i^2 x^4 - \alpha_i^2\beta_i^2 x^2)}{1-dx^2} = \frac{(x^2 - \beta_i^2)(1-d\alpha_i^2 x^2)}{1-dx^2}.$$

Similarly, we transform the denominator of the first coordinate (8)

$$1-(d\alpha_i\beta_i xy)^2 = 1-d^2\alpha_i^2\beta_i^2 x^2\frac{1-x^2}{1-dx^2} = \frac{1-dx^2 - d^2\alpha_i^2\beta_i^2 x^2 + d^2\alpha_i^2\beta_i^2 x^4}{1-dx^2} = \frac{1-d(\alpha_i^2+\beta_i^2)x^2 + d^2\alpha_i^2\beta_i^2 x^4}{1-dx^2} =$$

$$= \frac{(1-d\alpha_i^2 x^2)(1-d\beta_i^2 x^2)}{1-dx^2}.$$

After reducing the common factors, we obtain

$$\frac{(\alpha_i x)^2 - (\beta_i y)^2}{1-(d\alpha_i\beta_i xy)^2} = \frac{x^2 - \beta_i^2}{1-d\beta_i^2 x^2}$$

Similar calculations can be carried out with the second coordinate (8). As a result, the function (8) can be written in the equivalent form

$$\phi(x,y) = \left(\frac{x}{A^2}\prod_{i=1}^s\frac{x^2-\beta_i^2}{1-d\beta_i^2 x^2}, \frac{-y}{A^2}\prod_{i=1}^s\frac{x^2-\alpha_i^2}{1-d\alpha_i^2 x^2}\right), \qquad (9)$$

corresponding to the classical form (7). This form is given in [4] without proof. Its obvious advantage over (8) is simplicity and minimal computational complexity. Also, the degree of isogeny as the maximum degree of the polynomial $p(x)$ in (7) is immediately determined $l = 2s+1$.

Consider the example of a 3-isogeny of the complete supersingular Edwards curve.

**Example 1**. Let $E_d$ is a complete supersingular Edwards curve (2) at $p = 23, d = -1$. with the $j$-invariant $j = 12^3$ [6]. It has the order $N_E = 24$ and contains the points $(\pm 1, 0)$, $(0, \pm 1)$, $(\pm 2, \pm 2)$, $(\pm 3, \pm 6)$, $(\pm 6, \pm 3)$, $(\pm 9, \pm 10)$, $(\pm 10, \pm 9)$. Let's denote $P_1 = (3, 6)$, $P_2 = (6, 3)$ is the points of the $24^{\text{th}}$ order of the curve. $P_3 = (2, 2)$ is the point of the $8^{\text{th}}$ order, $P_4 = (9, 10)$ is the point of the $12^{\text{th}}$ order, $P_5 = (10, 9)$ is the point of the $6^{\text{th}}$ order and $Q = (-10, 9)$ is the point of the $3^{\text{rd}}$ order, and for any point $P = (x_1, y_1)$  $P^* = P + D_0 = (-x_1, -y_1)$. Let's note that the sum of the points of the $2^{\text{nd}}$ order and the $3^{\text{rd}}$ order gives the point of the $6^{\text{th}}$ order, that's why $x$ is coordinates of the points of the $3^{\text{rd}}$ and $6^{\text{th}}$ orders have reverse signs. Thus the kernel 3-isogeny contains the points $(1, 0)$, $(-10, \pm 9)$, i.e. $\alpha = -10$, $\beta = 9$, $A^2 = 8$, and according to Theorem 2 [3], the parameter of the isogenic curve $E_{d'}{'}$ is equal to $d' = -8^4 = -2$. This supersingular curve with the j-invariant $j = 3$, except for the points $O = (1, 0)$  $D_0 = (-1, 0)$, $\pm F_0 = (0, \pm 1)$, has the points of the first quadrant: $R_1 = (3, 5)$  $R_2 = (5, 3)$ is the points of the $24^{\text{th}}$ order of the curve. $R_3 = (7, 7)$ is the point of the $8^{\text{th}}$ order, $R_4 = (9, 11)$ is the point of the $6^{\text{th}}$ order, and $R_5 = (11, 9)$ is the point of the $12^{\text{th}}$ order and $Q' = (-9, 11)$ is the point of the $3^{\text{rd}}$ order. With the help of the function (9)

$$\phi(x, y) = \left( \frac{x}{8} \cdot \frac{x^2 - 9^2}{1 + 9^2 x^2}, \frac{-y}{8} \cdot \frac{x^2 - 10^2}{1 + 10^2 x^2} \right)$$

We compute:

$$\phi(\pm P_1 = (3, \pm 6)) = \left( \frac{3}{8} \cdot \frac{3^2 - 9^2}{1 + 9^2 \cdot 3^2}, \frac{\mp 6}{8} \frac{3^2 - 10^2}{1 + 10^2 \cdot 3^2} \right) = (-7, \pm 7) = \mp R_3^*, \ \ \phi(P_2 = (6, 3)) = (7, -7) = -R_3,$$

$$\phi(P_3 = (2, 2)) = (7, 7) = R_3,$$

$$\phi(P_4 = (9, \pm 10)) = (0, \mp 1) = \mp F_0,$$

$$\phi(P_5 = (10, 9)) = (-1, 0) = D_0,$$

$$\phi(\pm Q = (-10, \pm 9)) = (1, 0) = O, \phi(O = (1, 0)) = (1, 0) = O.$$

For the transformation of other points, we can use the property of the function (9) $\phi(\pm x, \pm y) = (\pm x', \pm y')$. There is a "3 in 1" mapping with compression $E_d$ 3 times (24 points of the curve $E_d$ are mapped in 8 points of the isogenic curve $E_d$). In particular, 8 points of the $24^{\text{th}}$ order together with 4 points of the $8^{\text{th}}$ order are mapped in 4 points of the $8^{\text{th}}$ order, 4 points of the $12^{\text{th}}$ order together with 2 points of the $4^{\text{th}}$ order are mapped in 2 points of the $4^{\text{th}}$ order, points of the $4^{\text{th}}$ order and the $2^{\text{nd}}$ order are mapped in a point of the $2^{\text{nd}}$ order and, at last, the points of the kernel are mapped in O. All transformations correspond to multiplying points by 3, similar to endomorphism $E \rightarrow 3E$.

Let's consider the example of 5-isogeny on the complete supersingular Edwards curve.

**Example 2.** At $p = 19$ and $d = -1$ the complete supersingular Edwards curve $E_{18}$ (2) has the order $N_E = 20$ and contains the points of the $1^{\text{st}}$ quadrant: $P_1 = (2, 8)$, $P_2 = (4, 6)$ is the points of the $20^{\text{th}}$ order of the curve, $P_3 = (8, 2)$ is the point of the $10^{\text{th}}$ order, and $Q_1 = (6, 4)$ is the point of the $5^{\text{th}}$ order. Then $Q_2 = 2Q_1 = (-8, -2)$ and the kernel of 5-isogeny contains the points $\{(1, 0), (6, \pm 4), (-8, \pm 2)\}$. Thus, $\alpha_1 = 6$, $\beta_1 = 4$, $\alpha_2 = -8$, $\beta_2 = -2$, $A = \alpha_1 \alpha_2 = 9$, $A^2 = 5$, and the parameter of the isogenic supersingular curve is $d' = -5^4 = -2$. It contains the points of the $1^{st}$ quadrant: $R_1 = (4,5)$, $R_2 = (7,9)$ is the points of the $20^{th}$ order of the curve $E_{d'}'$, $R_3 = (5,4)$, $R_4 = (9,7)$ is the points of the $10^{th}$ order. Here the 5-isogeny has the form (9)

$$\phi(x,y) = \left( \frac{x}{5} \cdot \frac{x^2-4^2}{1+4^2 x^2} \cdot \frac{x^2-2^2}{1+2^2 x^2}, \frac{-y}{5} \cdot \frac{x^2-6^2}{1+6^2 x^2} \cdot \frac{x^2-8^2}{1+8^2 x^2} \right).$$

Then

$$\phi(P_1 = (2,8)) = \left( \frac{2}{5} \cdot \frac{2^2-4^2}{1+4^2 2^2} \cdot \frac{2^2-2^2}{1+2^2 2^2}, \frac{-8}{5} \cdot \frac{2^2-6^2}{1+6^2 2^2} \cdot \frac{2^2-8^2}{1+8^2 2^2} \right) = (0,1) = F_0,$$

$$\phi(P_2 = (4,6)) = (0,-1) = -F_0,$$

$$\phi(P_3 = (8,2)) = (-1,0) = D_0,$$

$$\phi(\pm Q_1 = (6,\pm4)) = (1,0) = O, \quad \phi(O = (1,1)) = (1,0) = O.$$

This "5 in 1" mapping converts the subsets of 5 points of the curve $E_d$ into one of the 4 points of the isogenous curve of the $4^{th}$, $2^{nd}$ order, or the point $O$.

Here the function (9) acts similarly to the endomorphism $E_d \to 5E_d$, which reduces the orders of points of orders of magnitude of 5 by a factor of 5.

It is important to note that building isogeny of the composite order (for example, the $15^{th}$) is practically meaningless. It is enough to construct simpler 3-isogeny and 5-isogeny and use the property of their composition based on the homomorphism of the mapping $\phi$. As the sum of the subgroup of points of the $15^{th}$ order is the direct sum of subgroups of simple $3^{rd}$ and $5^{th}$ orders, i.e. $G_{15} = G_3 \oplus G_5$, then for the corresponding isogeny $\phi_{15} = \phi_3 \oplus \phi_5$ is true. This property drastically reduces the complexity of calculating the isogeny of composite degrees.

For constructing the isogenies of the degrees $l^k$, $l = 3,5,..,k = 2,3,..,m$, the obvious property of the group is used: any cyclic group of points $< G_k >$ of the order $l^k$ contains the subgroup of points $< G_{k-1} >$ of the order c $l^{k-1}$ and the subgroup $< G_1 >$ of the order $l$. The point of the order $l$ form $< G_k >$ is defined by the scalar product $l^{k-1}G_k$. Then, starting with the highest degree m, we can construct the sequence of $l$-isogenies $\{\phi_{m-i}\}$, compositions of which $\phi_{m-t} = \phi_{m-1} \oplus \phi_{m-2} \oplus .. \oplus \phi_{m-t+1}$ give $l^k$-isogeny at $t = m-k$. Such an algorithm, executed at most in m steps, has polynomial complexity.

The security of the SIDH algorithm [1] requires that the number of subgroups of the curve $E_d$ of the order $p+1 = 4 \cdot 3^m \cdot 5^n$ for protection against a quantum computer is more than 760 bits. To effectively solve this problem, the curves $E_d$ and $E_{d'}'$ are considered over the extension $F_{p^2}$ of the field $F_p$ (and the curve $E_d$ is set over a simple field).

The order of the supersingular curve over the extension $F_{p^2}$ is equal to $(p+1)^2$, in the corresponding proportion the number of subgroups of the curve (about 1.5 Kbit) increases. Every cyclic subgroup of the $n$ order of the supersingular curve over $F_p$ is transformed over the extension $F_{p^2}$ into a non-cyclic subgroup of the $n^2$ order, containing $(n+1)$ cyclic subgroups of the order $n$. Accordingly, the number of nuclei for 3-isogenies is 4, and for 5-isogenies is 6. Finding a generator of one such subgroup (or isogenic nucleus) is one of the challenges of PQC.

## 6. CALCULATION OF 3-ISOGENY IN PROJECTIVE COORDINATES

A promising solution to the problem of increasing the efficiency of isogeny computations is the transition to a single-coordinate isogeny $(X' : Z')$ [1, 11], whereas the second coordinate of the point with an accuracy of up to a sign is, if necessary, determined by the equation of the isogenous curve.

In this case the best results can be received by the use of isogeny of form (9).

For the first coordinate of 3-isogeny after substitution of $\beta^2 = (1-\alpha^2)/(1-d\alpha^2)$ we have:

$$\frac{X'}{Z'} = \frac{x}{\alpha^2} \cdot \frac{x^2 - \beta^2}{1 - d\beta^2 x^2} = \frac{x}{\alpha^2} \cdot \frac{x^2 - \dfrac{1-\alpha^2}{1-d\alpha^2}}{1 - dx^2 \cdot \dfrac{1-\alpha^2}{1-d\alpha^2}} = \frac{-x}{\alpha^2} \cdot \frac{x^2 + \alpha^2 - d\alpha^2 x^2 - 1}{d(x^2 + \alpha^2) - d\alpha^2 x^2 - 1}.$$

For the points of the kernel $\pm Q = (\alpha, \pm\beta)$ of the $3^{rd}$ order from the equation $2Q = -Q$ and the equation (5) it is easy to obtain the equation for the division polynomial $2\alpha + 1 - d\alpha^3(2+\alpha) = 0$, where $d = (2\alpha+1)/\alpha^3(2+\alpha)$ [11]. Substituting this value into the last equation, we arrive at a rational function

$$\frac{X'}{Z'} = x \cdot \frac{x^2 + \alpha^2 + 2\alpha}{x^2 + \alpha^2 + 2\alpha x^2}.$$

It is important that here the 3-isogene is determined only by the $x$-coordinates of the points P and Q and are independent of the parameter $d$.

In projective coordinates, after substitution $x \to \dfrac{X}{Z}$, $\alpha = \dfrac{X_1}{Z_1}$ we obtain

$$(X' : Z') = (X(X^2 Z_1^2 + X_1^2 Z^2 + 2X_1 Z_1 Z^2) : Z(X^2 Z_1^2 + X_1^2 Z^2 + 2X_1 Z_1 X^2)). \tag{10}$$

A similar expression was found in [11], in which instead of isogeny, defined by Theorem 2, Theorem 3 [4] was taken as the basis.

These theorems give different definitions of parameter $d'$ of isogenic curve $E_{d'}{}'$. Acoording to Theorem 2 [4],

$$d' = A^8 d^3, A = \alpha. \tag{11}$$

Defining here the parameter $d = (2\alpha+1)/\alpha^3(2+\alpha)$, in projective coordinates the equation (11) takes the form

$$d' = \frac{Z_1}{X_1} \cdot \frac{(2X_1 + Z_1)^3}{(2Z_1 + X_1)^3}. \tag{12}$$

In order to avoid inversion when calculating the parameter $d'$, it was proposed in [11] to use the projective coordinates of the isomorphic (2) curve

$$E_{C',D'}: \quad C'(x^2 + y^2) = C' + D'x^2 y^2, D' = d'C'.$$

Then, according to (12)

$$D' = Z_1(2X_1 + Z_1)^3 = (2X_1 Z_1 + Z_1^2)(4X_1^2 + Z_1^2 + 4X_1 Z_1), \quad (13)$$

$$C' = X_1(2Z_1 + X_1)^3 = (2X_1 Z_1 + X_1^2)(4Z_1^2 + X_1^2 + 4X_1 Z_1). \quad (14)$$

As $2X_1 Z_1 = (X_1 + Z_1)^2 - X_1^2 - Z_1^2$, the calculations by equations (13) and (14) have the cost $2M + 3S$.

The calculation of the coordinate (10) of the point $E_{d'}{}'$ of the isogenous curve can be performed using the equations [11]:

$$F = (X' + Z') = (X_1 Z + Z_1 X)^2 (X + Z), \quad (15)$$

$$G = (X' - Z') = (X_1 Z - Z_1 X)^2 (X - Z). \quad (16)$$

Then $2X' = F + G$, $2Z' = F - G$. $E_{d'}{}'$ calculations by equations (15) and (16) have the cost $4M + 2S$. The total cost of calculating 3-isogeny in projective coordinates is equal to $6M + 5S$.

## 7. 5-ISOGENY COMPUTATION IN PROJECTIVE COORDINATES

For the first coordinate of 5-isogene (9) after substitution $\beta_{1,2}^2 = (1 - \alpha_{1,2}^2)/(1 - d\alpha_{1,2}^2)$ we obtain:

$$\frac{X'}{Z'} = \frac{x}{(\alpha_1 \alpha_2)^2} \cdot \frac{x^2 + \alpha_1^2 - d\alpha_1^2 x^2 - 1}{d(x^2 + \alpha_1^2) - d\alpha_1^2 x^2 - 1} \cdot \frac{x^2 + \alpha_2^2 - d\alpha_2^2 x^2 - 1}{d(x^2 + \alpha_2^2) - d\alpha_2^2 x^2 - 1}.$$

For this case, the division polynomial for points of the 5th order has the degree 12, and it includes coefficients with parameters $d^m, m \le 3$. Using the division polynomial here does not give the same effect as for 3-isogeny.

In projective coordinates after substitution $x \to \dfrac{X}{Z}$, $\alpha_{1,2} \to \dfrac{X_{1,2}}{Z_{1,2}}$ we have

$$\frac{X'}{Z'} = \frac{X(Z_1 Z_2)^2}{Z(X_1 X_2)^2} \cdot \frac{(XZ_1)^2 + (X_1 Z)^2 - d(X_1 X)^2 - (Z_1 Z)^2}{d((XZ_1)^2 + (X_1 Z)^2) - d(X_1 X)^2 - (Z_1 Z)^2} \cdot \frac{(XZ_2)^2 + (X_2 Z)^2 - d(X_2 X)^2 - (Z_2 Z)^2}{d((XZ_2)^2 + (X_2 Z)^2) - d(X_2 X)^2 - (Z_2 Z)^2}.$$

Accordingly,

$$X' = XZ_1^2 Z_2^2 [X^2 (Z_1^2 - dX_1^2) + Z^2 (X_1^2 - Z_1^2)][X^2 (Z_2^2 - dX_2^2) + Z^2 (X_2^2 - Z_2^2)], \quad (17)$$

$$Z' = ZX_1^2 X_2^2 [dX^2 (Z_1^2 - X_1^2) + Z^2 (dX_1^2 - Z_1^2)][dX^2 (Z_2^2 - X_2^2) + Z^2 (dX_2^2 - Z_2^2)]. \quad (18)$$

The calculations according to equation (17) and (18) require $19M + 6S$. The parameter $d'$ of isogenous curve is defined as

$$d' = A^8 d^5, A = \alpha_1 \alpha_2.$$

The parameters of the isomorphic curve $E_{C',D'}$ are at the same time equal to

$$D' = (X_1^2 X_2^2)^4 \cdot (d^2)^2 \cdot d, \tag{19}$$

$$C' = (Z_1^2 Z_2^2)^4 . \tag{20}$$

The calculations by equations (19) and (20) have the cost $2M + 6S$. General cost of 5-isogene computing is $21M + 12S$.

## 8. ALGORITHMS FOR 3- AND 5-ISOGENOUS EDWARDS CURVES

The calculation of the 3- and 5-isogenes of the Edwards curves (2) according to equations (13)–(20) and the calculation of the parameter $d' = A^8 d^l$ of the isogeous curve is performed using the following algorithms with the cost $6M + 5S$ and $21M + 12S$ respectively.

---

Entry: point $P = (X : Z)$ and point of 3<sup>rd</sup> order $Q_1 = (X_1 : Z_1)$, of kernel of curve $E_d$ with parameter $d$

1. $s_1 \leftarrow X_1^2$
2. $s_2 \leftarrow Z_1^2$
3. $t_1 \leftarrow (X + Z_1)^2 - s_0 - s_2$
4. $t_2 \leftarrow t_1 + s_1$
5. $t_3 \leftarrow t_1 + s_2$
6. $t_4 \leftarrow 2t_1$
7. $t_5 \leftarrow 4s_1 + s_2 + t_4$
8. $t_6 \leftarrow 4s_2 + s_1 + t_4$
9. $D' \leftarrow t_3 \cdot t_5$
10. $C' \leftarrow t_2 \cdot t_6$
11. $u_1 \leftarrow X_1 \cdot Z$
12. $u_2 \leftarrow X \cdot Z_1$
13. $u_3 \leftarrow (u_1 + u_2)^2$
14. $u_4 \leftarrow (u_1 - u_2)^2$
15. $F \leftarrow (X + Z) \cdot u_3$
16. $G \leftarrow (X - Z) \cdot u_4$
17. $2X' \leftarrow F + G$
18. $2Z' \leftarrow F - G$

Exit: point of curve $E_{d'}^r$ $P' = (X' : Z')$ and parameter $(D' : C')$ of isogenic curve $C'E_{d'}'$

---

*Algorythm 1. Calculation of 3-isogene Edwards curve*

Entry: point $P = (X : Z)$ and point of $5^{\text{th}}$ order $Q_1 = (X_1 : Z_1)$, $Q_2 = (X_2 : Z_2)$ of kernel of curve $E_d$ with parameter $d$

1. $s_0 \leftarrow X^2$
2. $s_1 \leftarrow X_1^{\ 2}$
3. $s_2 \leftarrow X_2^{\ 2}$
4. $s_3 \leftarrow Z^2$
5. $s_4 \leftarrow Z_1^{\ 2}$
6. $s_5 \leftarrow Z_2^{\ 2}$
7. $t_0 \leftarrow d \cdot s_0$
8. $t_1 \leftarrow d \cdot s_1$
9. $t_2 \leftarrow d \cdot s_2$
10. $G_1 \leftarrow s_1 - s_4$
11. $F_1 \leftarrow t_1 - s_4$
12. $G_2 \leftarrow s_2 - s_5$
13. $F_2 \leftarrow t_2 - s_5$
14. $H_1 \leftarrow -s_o \cdot F_1 + s_3 \cdot G_1$
15. $I_1 \leftarrow -t_o \cdot G_1 + s_3 \cdot F_1$
16. $H_2 \leftarrow -s_o \cdot F_2 + s_3 \cdot G_2$
17. $I_2 \leftarrow -t_o \cdot G_2 + s_3 \cdot F_2$
18. $L_1 \leftarrow s_1 \cdot s_2$
19. $L_2 \leftarrow s_4 \cdot s_5$
20. $X' \leftarrow X \cdot L_2$
21. $X' \leftarrow X' \cdot H_1$
22. $X' \leftarrow X' \cdot H_2$
23. $Z' \leftarrow Z \cdot L_1$
24. $Z' \leftarrow Z' \cdot I_1$
25. $Z' \leftarrow Z' \cdot I_2$
26. $D \leftarrow d^2$
27. $D \leftarrow D^2$
28. $D \leftarrow D \cdot d$
29. $L \leftarrow L_1^{\ 2}$
30. $L \leftarrow L^2$
31. $D' \leftarrow L \cdot D$
32. $C \leftarrow L_2^{\ 2}$
33. $C' \leftarrow C^2$

Exit: point of curve $E_{d'}'$ $P' = (X' : Z')$ and parameter $(D' : C')$ of isogenic curve $C'E_{d'}'$

*Algorithm 2. Calculation of 5-isogeny Edwards curve*

These algorithms are distinguished by the greatest simplicity and low cost of computing among the known ones. In contrast to the 3-isogeny calculation algorithm given in [11] and instead of (8) we use the simpler expression (9) for the function $\phi(x, y)$ together with the simpler equation for the parameter $d' = A^8 d^l$.

In fact, when calculating 3-isogeny, we use an algorithm close to that proposed in [11], with the same effectiveness $6M + 5S$. Our algorithm for computing 5-isogeny is almost three times slower than for 3-isogeny, and probably has reserves to increase efficiency.

## 9. REQUIREMENTS FOR CRYPTOSYSTEM PARAMETERS

The search for a suitable value of the characteristic of the field $p$ in the SIDH problem using 3- and 5-isogeny of Edwards curves must meet several necessary conditions.

**Statement 1**. 3- and 5-isogenies exist for supersingular complete and quadratic Edwards curves $E_d$, respectively, at $p \equiv -1 \bmod 60$ and $p \equiv -1 \bmod 120$.

**Proof**. Points of the $3^{rd}$ and $5^{th}$ orders exist on the complete supersingular Edwards curve of $p + 1 = 4 \cdot 3^m \cdot 5^n$ order under the conditions that $p \equiv -1 \bmod 4$, $p \equiv -1 \bmod 3$ and $p \equiv -1 \bmod 5$, and which come down to one condition $p \equiv -1 \bmod 60$.

The minimum even cofactor of the order $N_E$ of the quadratic Edwards curve is the number 8 [7], at the same time at $p + 1 = 8 \cdot 3^m \cdot 5^n$ the condition $p \equiv -1 \bmod 120$ is true.

**Statement 2**. For odd $l = 2s + 1$ of $l$-isogeny of P points of odd order of the curve there are points of odd order.

**Proof**. The Edwards curve $E_d$ of the order $N_E = 2^c \cdot n, c \geq 2$, contains the points $P$ of the odd order $n = l \cdot m$. Thus $l$-isogene and isogenous curve $E'$ of the same order $N_E$ exist. $l$-isogeny is a homomorphism that compresses $l$ times the points $< P >$ into a subgroup of points of odd order of the curve $E'$. This subgroup does not contain any ponts of even order. At $m = 1$ $n$-isogeny maps all points $< P >$ into neutral element $O$ of the order1.

**Statement 3**. At $p \equiv 1 \bmod 4$ **supersingular Edwards curves do not exist.**

**Proof**. At $p \equiv 1 \bmod 4$ the order of **supersingular curve** is $p + 1 \equiv 2 \bmod 4$, at the same time for any Edwards curve the number 4 divides the order of the curve.

The value of the module $p$ is determined by the security requirements. In the product $3^m \cdot 5^n$ both factors have the same order at $3^m \approx 5^n$, then $m \approx 1.465n$. This balances the number of corresponding cyclic subgroups.

128-bit quantum security with complexity estimate $\sqrt[6]{p}$ ( instead of $\sqrt[4]{p}$ for a regular computer) is provided with the length of module $\log_2 p = 6 \cdot 128 = 768$ bit. In the field $F_{p^2}$ each coordinate of the point has the length $2 \log_2 p = 1536$ bit. The key length estimate in the SIDH system is $6 \cdot \log_2 p = 6 \cdot 768 = 4608$ bit. 256-bit quantum security level doubles all of these estimates.

Below are 3 field module values of the field $p$ found by brute force with a length of about 768 bits for the implementation of the SIDH algorithm on the 3- and 5-isogeny of complete Edwards curves (see Table 1).

*Table 1*

**Automatic search results by SIDH algorithm**

| # | m | n | $p = 4 \times 3^m \times 5^n - 1$ | log p, bit |
|---|---|---|---|---|
| 1 | 238 | 165 | 0×50f6d0ab1dad4fb9048ca2e5357e7fa140806f49f72b711a651962fd24d6ae3095 3eeb9cafca76f39eae708b2bfa6926d7df2937074b004fa4d966e8ecd7469bc771d4 dd084b5a9f358a2c83e4f67398f1b7972610af76087956accd41b0c33 | 763 |
| 2 | 243 | 168 | 0×25869530ff4e3ece49cacad3ea2e345995ec4714b12e4378f2d1a730421dfc5606 7c5ca5ec3dffe7e410ebab910f1cd27fd7af9340425411e9f0bf417f1dbafadd8d935f be0324ed80899da7d593f60de8304e6f2585c2dde7751b31562d544edeb | 778 |
| 3 | 247 | 156 | 0×d0e0e81c7cf2831a189cf43da28062552d4a98e390e7b3f3bb8bd34b91e364d78 49480255df7222b93e45fe7640850a6e60e1afd64a07ee55f821e7009ec557cfbd9a bca5dd1b758d06ec0939ca37cc685f937196f3bd26aa01ae966c35eb | 756 |

## 10. CONCLUSION

Thus, the use of 3- and 5-isogeny of Edwards curves for points of odd order with a fixed resistance to attacks by a quantum computer will allow bypassing the problems of singular points inherent to 2-isogenies of these curves. Estimates of the complexity of computing the 3- and 5-isogeny of Edwards curves, comparable to w the complexity of group operations, allow us to implement the fastest post-quantum cryptography algorithms. Experimental estimates of the computational efficiency of these isogenies in the implementation of the SIDH algorithm are planned to be considered in the next paper.

**Бессалов Анатолій Володимирович**
д. т. н., професор
Київський університет імені Бориса Грінченка, Київ, Україна
ORCID: 0000-0002-6967-5001
*a.bessalov@kubg.edu.ua*

**Грубіян Євген Олександрович**
магістр
Національний технічний університет України «Київський політехнічний інститут імені Ігоря Сікорського», Київ, Україна
ORCID: 0000-0002-6967-5001
*grubian.euhen@gmail.com*

**Соколов Володимир Юрійович**
к. т. н., доцент кафедри інформаційної та кібернетичної безпеки
Київський університет імені Бориса Грінченка, Київ, Україна
ORCID: 0000-0002-9349-7946
*v.sokolov@kubg.edu.ua*

**Складанний Павло Миколайович**
старший викладач кафедри інформаційної та кібернетичної безпеки
Київський університет імені Бориса Грінченка, Київ, Україна
ORCID: 0000-0002-7775-6039
*p.skladannyi@kubg.edu.ua*

## 3- І 5-ІЗОГЕНІЇ СУПЕРСІНГУЛЯРНИХ КРИВИХ ЕДВАРДСА

**Анотація.** Дан аналіз властивостей і умов існування 3- і 5-ізогеній повних і квадратичних суперсінгулярних кривих Едвардса. Для завдання інкапсуляції ключів на основі алгоритму SIDH запропоновано використовувати ізогенії мінімальних непарних ступенів 3 і 5, що дозволяє обійти проблему особливих точок 2-го і 4-го порядків, характерну для 2-ізогеній. Наведено огляд основних властивостей класів повних, квадратичних і скручених кривих Едвардса над простим полем. Формули для ізогеній непарних ступенів приведені до вигляду, адаптованому до кривих в формі Вейєрштрасса. Для цього використовується модифікований закон складання точок кривої в узагальненій формі Едвардса, який зберігає горизонтальну симетрію зворотних точок кривої. Наведені приклади обчислення 3- і 5-ізогенна повних суперсінгулярних кривих Едвардса над малими простими полями і обговорюються властивості композиції ізогеній для їх обчислення з ядрами високих порядків. Отримано формули верхніх оцінок складності обчислень ізогеній непарних ступенів з 3 і 5 в класах повних і квадратичних кривих Едвардса в проективних координатах побудовано алгоритми обчислення 3- і 5-ізогеній кривих Едвардса зі складністю $6M + 4S$ і $12M + 5S$ відповідно. Знайдено умови існування суперсінгулярних повних і квадратичних кривих Едвардса порядку $4 \cdot 3^m \cdot 5^n$ і $8 \cdot 3^m \cdot 5^n$. Визначено деякі параметри криптосистеми при реалізації алгоритму SIDH на рівні квантової безпеки 128 біт.

**Ключові слова:** крива в узагальненій формі Едвардса, повна крива Едвардса, скручена крива Едвардса, квадратична крива Едвардса, порядок кривої, порядок точки, ізоморфізм, ізогенія, ступінь ізогенії, ядро ізогенії, квадратичний відрахування, квадратичний невирахування.

## СПИСОК ВИКОРИСТАНИХ ДЖЕРЕЛ